\documentclass{cta-author}

{}
{}
{}
\usepackage{amsmath}
\usepackage{graphicx}
\usepackage{subfigure}
\usepackage{graphicx}
\usepackage{epstopdf}
\usepackage{verbatim}
\usepackage{color}
\usepackage{float}
\begin{document}

\supertitle{Research Article}

\title{Secrecy Outage Probability Analysis for RIS-Assisted NOMA Systems}

\author{\au{Liang Yang$^{1\corr}$}, \au{Yongjie Yuan$^{1}$}}

\address{\add{1}{College of Computer Science and Electronic Engineering, Hunan University, Changsha, China}
\email{liangy@hnu.edu.cn}}

\begin{abstract}
In this paper, the physical layer security (PLS) for a novel
reconfigurable intelligent surface (RIS)-assisted non-orthogonal multiple access (NOMA)
system in a multi-user scenario is investigated, where we consider the worst case that the
eavesdropper also utilizes the advantage of the RISs. More specifically, we derive analytical
results for the secrecy outage probability (SOP). From the
numerical results, we observe that the use of RISs can
improve the secrecy performance compared to traditional NOMA systems. However, for the worst case
that the received signals at the eavesdropper comes from the RISs and source,
increasing the number of intelligent elements on the RIS has a negative impact on the secrecy performance. At high SNRs, the system's SOP tends to a constant. Finally, the secrecy performance can be improved through the group selection.
\end{abstract}

\maketitle

\section{Introduction}\label{sec1}
5G has been commercialized in 2020, and
non-orthogonal multiple access (NOMA) plays a key role in this.
NOMA has been used for many scenarios to solve the problems caused
by the explosive growth of the number of mobile terminals [1].
Unlike the traditional orthogonal multiple access (OMA) system
structure, power-domain NOMA serves more users at the same time
and frequency based on the power allocation of the transmitted
signals. Consequently, NOMA can enhance the communication quality
of users in poor channel conditions [2]. Recently, a
multiple-input and multiple-output (MIMO)-NOMA system which uses
max-min transmit antenna selection strategy was proposed in [3].
In [4], the authors proposed a new hybrid decode-and-forward (DF)
and amplify-and-forward (AF) transmission mode for a
multiple-relay NOMA system. A NOMA system under Rician fading
channels was studied in [5] and  expressions of
the average achievable rate were derived.

In wireless communication systems, the signals are broadcast so
that physical layer security (PLS) has become a hot issue. The
analysis of secrecy performance in various wireless systems is
studied in the literature, such as dual-hop RF/free-space optical
(FSO) systems [6] and transmit antenna selection (TAS)/maximal
ratio combining (MRC) systems [7]. Recently, many works about PLS
for NOMA systems have been considered. For instance, in [8], PLS
for cognitive radio inspired NOMA networks was investigated. PLS
in a multiuser visible light communication (VLC) system with NOMA
was considered in [9]. In [10], PLS for cooperative NOMA systems
was investigated, where both AF and DF were considered.

Recently, a new material called reconfigurable intelligent
surfaces (RIS) has been proposed. RISs have a large number of
application scenarios in wireless communication, and even change
the traditional communication structure [11]. So far, there are many
works based on RISs have been reported in [12-22]. For example,
a mixed dual-hop FSO-RF system through the RIS was proposed in [12]. An RIS-assisted dual-hop UAV communication system was proposed in [13]. The authors in [14] quantitatively analyzed the coverage for an RIS-aided communication system. An RIS-aided downlink multi-user communication system was investigated in [15].
The authors in [16] proposed a deep learning method for deploying RISs in an indoor environment.
Moreover, an important application of RISs is to combine with NOMA to
further improve communication quality. For instance,
the authors in [17] proposed an RIS-empowered NOMA network to introduce
desirable channel gain differences by adjusting the phase shifts at the RISs.
In [18], the authors conceived a system for serving paired power-domain multiple NOMA
users by designing the passive beamforming weights at the RISs.
The authors in [19] proved that NOMA can achieve the capacity region when the channels
are quasi-degraded by using the RISs. In [20], the authors proposed a theoretical performance comparison
between NOMA and OMA in the RIS-assisted downlink communication.
The authors in [21] derived the bit error rate (BER) performance of
the RIS-assisted power domain NOMA system. In [22],
the authors studied both downlink and uplink RIS-aided NOMA and OMA networks.
However, considering the PLS for the RIS-aided NOMA
system is still not reported in the literature. Therefore, this is
the main innovation of this work.

In this paper, we propose an RIS-assisted multi-user NOMA system. In
particular, we assume that an eavesdropper in the considered
network can receive signals from the RISs and source to affect the legitimate
users. Based on this assumption, we intend to investigate whether the
RIS always improves the secrecy performance. In particular, we derive analytical
expressions for the secrecy outage probability (SOP). Also, the
asymptotic SOP analysis at high signal-to-noise ratio (SNR)
condition is provided. Finally, some numerical results are
presented to verify our analysis and investigate the effects of
the number of reflecting surfaces in the RIS on the system secrecy
performance.

\section{System and channel models}\label{sec2}
As shown in Figure 1, consider an RIS-assisted NOMA system which
includes a source (S), $M$ RISs, $M$ groups of $U=\{({\rm UE}_{11},{\rm UE}_{21}), ... , ({\rm UE}_{m1},{\rm UE}_{m2}), ... ,({\rm UE}_{M1},{\rm UE}_{M2})\}$ NOMA users, and an eavesdropper (E), where $M$ near users are close to S, while $M$ far users have a long distance from S. Therefore, similar to [22] and [23], we utilize $M$ RISs to increase the signal coverage to improve the $M$ far users' communication quality, while the near users directly communicate with S. We assume that $M$ RISs have the same $N$ reflecting elements. Furthermore, we assume the worst case that E can utilize the advantage of the RIS. Finally, we suppose that the channels in
this system suffer from Rayleigh fading independently.
\begin{figure}[h]
\centerline{\includegraphics[width=8cm,height=5cm]{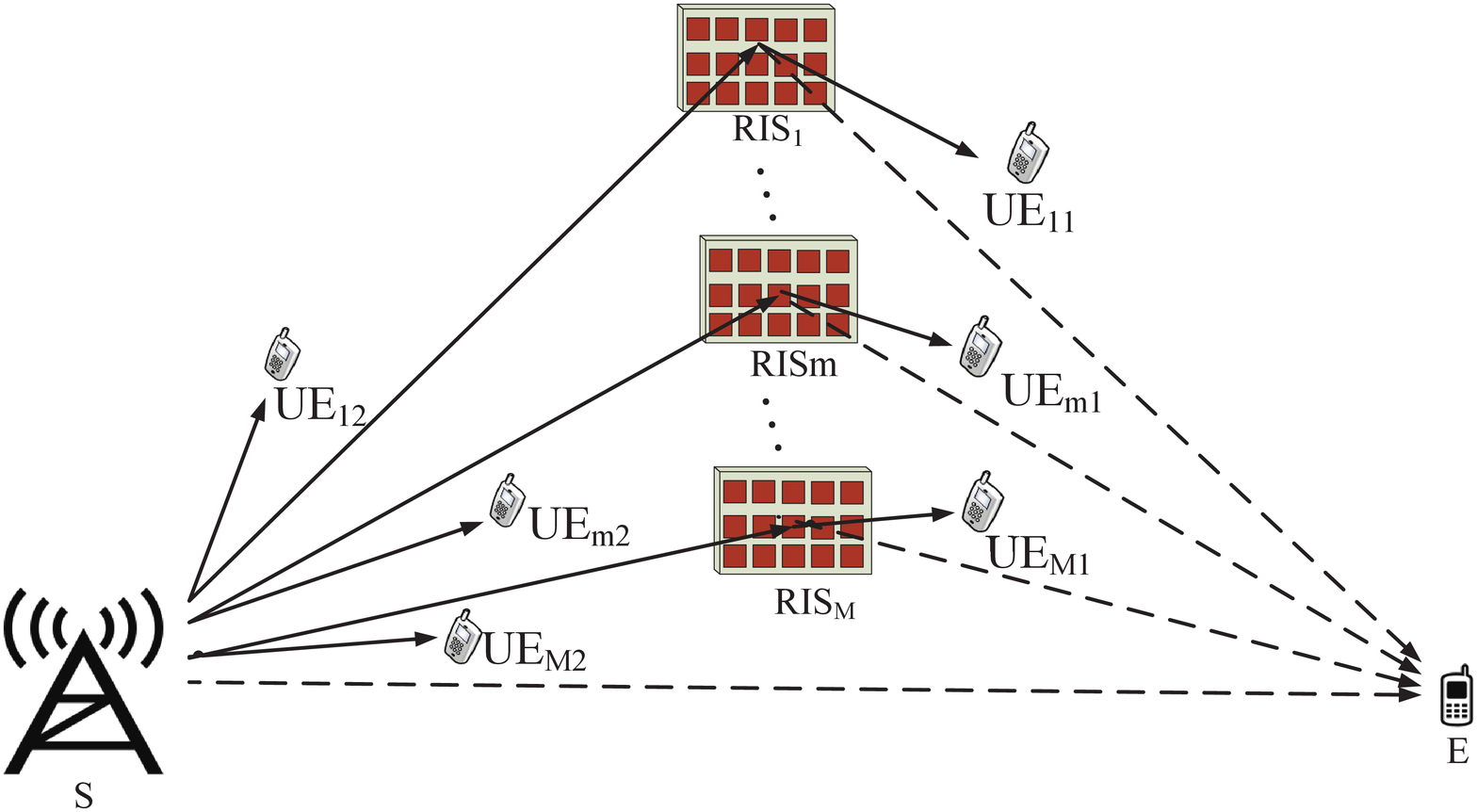}}
\caption{System diagram of considered system.\label{fig1}}
\end{figure}
According to the NOMA protocol, we need to distribute the total transmit power to the $2M$ NOMA users concurrently,
but it is not preferable to group all the users in a NOMA system in practice [24]. Therefore, we choose one near user and one far user to constitute a NOMA group, and then in every group we use one RIS to improve the far user's received SNR.
Thus, $2M$ users are divided into $M$ groups and only one group can be selected to communicate according to the criterion described later.

In particular, we assume that the far user ${\rm UE}_{m1}$ with poor channel gains is defined as the weak user and the near user ${\rm UE}_{m2}$ with good channel gains is defined as the strong user ($m=1,2,...,M$). In order to enhance ${\rm UE}_{m1}$'s communication quality, we set that $c_{m1}
\geq c_{m2}$, and let them satisfy $c_{m1}^2+c_{m2}^2=1$ [10], where $c_n$ is the power distribution coefficient ($n=m1,m2$). First, the mixed signal, $(c_{m1}s_{m1}+c_{m2}s_{m2})$, is broadcast from S to
the $m$th RIS (${\rm RIS}_{m}$) and the near user, where $s_n$ is the unit signal needed by user
$n$. Then, ${\rm RIS}_{m}$ passively reflects the signals to ${\rm UE}_{m1}$. Thus, the received signal by the far user ${\rm UE}_{m1}$ can be written as
\begin{equation}
y_{m1}=\left[\sum_{i=1}^Nh_{mi}e^{j\phi _{mi}}g_{m1i}\right](c_{m1}s_{m1}+c_{m2}s_{m2})+w_{m1},
\end{equation}
where $h_{mi}$ and $g_{m1i}$ are the channel gains for the S-${\rm RIS}_{m}$ and ${\rm RIS}_{m}$-${\rm UE}_{m1}$ links. In (1), $\phi_{mi}$ is the adjustable
phase produced by the $i$th reflecting element of ${\rm RIS}_{m}$
($i=1,2,...,N$). Let $h_{mi}=d_{SR_{m}}^{-\chi/2}\alpha
_{mi}e^{-j\theta_{mi}}$ and $g_{m1i}=d_{R_{m}U_{m1}}^{-\chi/2}\beta
_{m1i}e^{-j\varepsilon_{m1i}}$, where $d_{SR_{m}}$ and $d_{R_{m}U_{m1}}$ are the distances for the S-${\rm RIS}_{m}$ and ${\rm RIS}_{m}$-${\rm UE}_{m1}$ links,
$\chi$ denotes the path loss coefficient,
$\alpha _{mi}$ and $\beta _{m1i}$ denote the
channels' amplitudes, $\theta_{mi}$ and $\varepsilon_{m1i}$ are the phases of the fading channels. Similar to [11],
we assume that ${\rm RIS}_{m}$ has perfect knowledge of the channels
phases of $h_{mi}$ and $g_{m1i}$.

For the near users, they receive mixed signals from S directly. Thus, the received signals by the near user ${\rm UE}_{m2}$ can be written as
\begin{equation}
y_{m2}=h_{m2}d_{SU_{m2}}^{-\chi/2}(c_{m1}s_{m1}+c_{m2}s_{m2})\sqrt{E_{s}}+w_{m2},
\end{equation}
where $d_{SU_{m2}}$ is the distance of the S-${\rm UE}_{m2}$ link and $E_s$ is the average transmitted energy per symbol. In (1) and (2), $w_{m1}$
and $w_{m2}\sim\mathcal C\mathcal N(0,N_0)$ are the additive white
Gaussian noise (AWGN) samples.

Since E receives the same signals from ${\rm RIS}_{m}$ and S, the received signal at E can be expressed as
\begin{equation}
y_{Em}{=}\left[\frac{h_{E}}{d_{SE}^{\chi/2}}{+}\sum_{i=1}^Nh_{mi}e^{j\phi _{mi}}g_{Emi}\right](c_{m1}s_{m1}{+}c_{m2}s_{m2}){+}{w_E}_m,
\end{equation}
where $g_{Emi}=d_{R_{m}E}^{-\chi/2}\beta _{Emi}e^{-j\delta_{mi}}$,
$d_{R_{m}E}$ and $d_{SE}$ denote the distance for the ${\rm RIS}_{m}$-E and S-E links, $\beta _{Emi}$ and $\delta_{mi}$ are the amplitude and phase of the fading channel, and ${w_E}_m$ is the AWGN
sample with variance $N_E$.

According to [25], in NOMA systems, we can use successive
interference cancellation (SIC) technology to decode the signals
of different users. For the weak user ${\rm UE}_{m1}$, it has poor channel gains. ${\rm UE}_{m1}$ normally decodes its
own signal, but it has no power to remove the signal of ${\rm UE}_{m2}$ from
the mixed signals. Thus, ${\rm UE}_{m1}$ suffers from slight extra
interference from ${\rm UE}_{m2}$. Hence, the instantaneous
signal-to-interference-noise ratio (SINR) for ${\rm UE}_{m1}$ can be expressed
as
\begin{equation}
\gamma _{m1}=\frac{\left|\sum_{i=1}^N\alpha_{mi}\beta_{m1i}e^{j(\phi _{mi}-\theta_{mi}-\varepsilon_{m1i})}\right|^2c_{m1}^2Es}{\left|\sum_{i=1}^N\alpha_{mi}\beta_{m1i}e^{j(\phi_{mi}-\theta_{mi}-\varepsilon_{m1i})}\right|^2c_{m2}^2Es+N_{0}\varpi}.
\end{equation}
where $\varpi = d_{SR_{m}}^{\chi}d_{R_{m}U_{m1}}^{\chi}$.

For the legal far users, we assume that their channel state
information (CSI) is known to the RIS. Like [11], the ${\rm RIS}_m$ can use the phase
shifting to maximize $\gamma_{m1}$ when $\phi_{mi}=\theta_{mi}+\varepsilon_{m1i}$. Therefore, the maximized
$\gamma_{m1}$ can be written as
\begin{equation}
\begin{aligned}
\gamma_{m1}&=\frac{\left(\sum_{i=1}^N\alpha_{mi}\beta_{m1i}\right)^2c_{m1}^2Es}{\left(\sum_{i=1}^N\alpha_{mi}\beta_{m1i}\right)^2c_{m2}^2Es+N_{0}\varpi}\\
&=\frac{A^2c_{m1}^2Es}{A^2c_{m2}^2Es+N_{0}\varpi},
\end{aligned}
\end{equation}
where $A=\sum_{i=1}^N\alpha_{mi}\beta_{m1i}$.

According to the center limit theorem (CLT), $A$ is a
Gaussian distributed random variable i.e., $A\sim\mathcal
N\left(\frac{N\pi}{4},N\left(1-\frac{\pi^2}{16}\right)\right)$.
Therefore, $A^2$ is a non-central chi-square random variable with
one degree of freedom [11]. From [26], the probability density
function (PDF) of $A^2$  can be written as
\begin{equation}
f_{A^2}(y)=\frac{1}{2\sigma^2}\left(\frac{y}{\lambda}\right)^{-\frac{1}{4}}{\rm exp}\left(-\frac{y+\lambda}{2\sigma^2}\right){\rm I}_{-\frac{1}{2}}\left(\frac{\sqrt{y\lambda}}{\sigma^2}\right),
\end{equation}
where ${\rm I_a(\cdot)}$ is the first order modified Bessel function, $\lambda=\left(\frac{N\pi}{4}\right)^2$, and $\sigma^2=N\left(1-\frac{\pi^2}{16}\right)$.

On the other hand, the strong user ${\rm UE}_{m2}$ also receives the mixed
signal, and it has bigger channel gain than ${\rm UE}_{m1}$ so that it
can get more energy. Therefore, ${\rm UE}_{m2}$ can decode the signal of ${\rm UE}_{m1}$
first, and then use the complete mixed signal to reduce the
interference signal of ${\rm UE}_{m1}$. Through this process, it can get a
clean signal of its own and then decode it. Therefore, we have
\begin{equation}
\gamma _{m2}=\frac{\left|h_{m2}\right|^2c_{m2}^2Es}{N_{0}d_{SU_{m2}}^{\chi}}.
\end{equation}

For the eavesdropper, we assume that the CSI of E is not known and ${\rm RIS}_{m}$ can not maximize the eavesdropper's SNR
to protect the communication of legitimate users. Similar to [25], we
assume that E has the multiuser detection ability, and it can use the
parallel interference cancellation (PIC) technology to intercept
the different users' signal. Then, the received SNR at E is
\begin{equation}
\begin{aligned}
\gamma _{En}&=\frac{\left|h_{E}d_{SE}^{-\chi/2}{+}\frac{\sum_{i=1}^N\alpha_{mi}\beta_{Emi}e^{j(\phi _{mi}-\theta_{mi}-\delta_{mi})}}{d_{SR_{m}}^{\chi/2}d_{R_{m}E}^{\chi/2}}\right|^2c_{n}^2Es}{N_{E}}\\
&=B^2c_{n}^2,
\end{aligned}
\end{equation}
where $n = (m1, m2)$ and $B^2$ follows the exponential distribution with parameter
$\lambda_E$ [27] and its PDF can be written as
\begin{equation}
f_{B^2}(y)=\frac{1}{\lambda_E}{\rm exp}\left(-\frac{y}{\lambda_E}\right).
\end{equation}
where $\lambda_E=N\overline{\gamma}_{SR_{m}E}+\overline{\gamma}_{SE}$, $\overline{\gamma}_{SR_{m}E}=\frac{E_{s}}{N_{E}d_{SR_{m}}^{\chi}d_{R_{m}E}^{\chi}}$ and $\overline{\gamma}_{SE}=\frac{E_{s}}{N_{E}d_{SE}^{\chi}}$ are the average SNRs.

Finally, the secrecy rates of the group $m$ for two
paired users can be expressed as
\begin{equation}
C_{m1}=\lceil\log_2(1+\gamma_{m1})-\log_2(1+\gamma_{E_{m1}})\rceil^{+},
\end{equation}
\begin{equation}
C_{m2}=\lceil\log_2(1+\gamma_{m2})-\log_2(1+\gamma_{E_{m2}})\rceil^{+},
\end{equation}
where $\lceil x \rceil^{+}=\max\{x,0\}$.

To obtain the best secrecy performance, it is optimal to select the group with the maximum achievable secrecy rate as the intended pairing mechanism. For arbitrary group $m$, when either $C_{m1}$ or $C_{m2}$ is
lower than the legal users' target rate, system outage appears. Therefore, the group selection policy is given by
\begin{equation}
m^\ast = {\rm arg} \max \limits_{{\rm UE}_{m1},{\rm UE}_{m2}\in U}({\rm min}\{C_{m1},C_{m2}\}<R).
\end{equation}

\section{Secrecy Performance Analysis}\label{sec3}
In this section, we present the calculation of the SOP. To get
more insights, an asymptotic SOP analysis is also presented.
\subsection{SOP analysis}\label{subsec3.1}
For $M$ paired groups, we assume that different groups are allocated with orthogonal bandwidth resources and have independent and identical distributions. With the group selection introduced in (12), the system achieves the best secrecy performance. Then, the system SOP in a multi-user scenario can be evaluated by
\begin{equation}
\begin{aligned}
{\rm SOP}&={\rm P_r}({\min\{C_{m^{\ast}1},C_{m^{\ast}2}\}}<R)\\
&=[{\rm P_r}(\min\{C_{m1},C_{m2}\}<R)]^{M}.
\end{aligned}
\end{equation}

Therefore, we need first to calculate ${\rm P_r}(\min\{C_{m1},C_{m2}\}<R)$. For the paired casual group $m$, when either $C_{m1}$ or $C_{m2}$ is
less than the legal users' target rate, this group outage appears.
Thus, the SOP of arbitrary group $m$ can be calculated as
\begin{align}
{\rm P_r}&(\min\{C_{m1},C_{m2}\}<R)=1-{\rm P_r}(C_{m1}>R,C_{m2}>R)\nonumber\\
&=1-{\rm P_r}\left(\frac{1+\gamma_{m1}}{1+\gamma_{E_{m1}}}>C_{th},\frac{1+\gamma_{m2}}{1+\gamma_{E_{m2}}}>C_{th}\right)=1-{\rm P_{r1}},
\end{align}
where $C_{th}=2^{R}$. From (4), (7), (14), it is very difficult to obtain the exact analysis.
Consequently, for tractable analysis, we consider a high SNR case and get
the upper bound $\gamma_{m1}<\frac{c_{m1}^2}{c_{m2}^2}$.  Later in numerical results,
we can see this upper bound is very tight to the exact simulation results.
Then, ${\rm P_{r1}}$ can be expressed
as
\begin{equation}
\begin{aligned}
{\rm P_{r1}}{<}& \ {\rm P_r}\left(\frac{1+\frac{c_{m1}^2}{c_{m2}^2}}{1+B^2c_{m1}^2}>C_{th},
\frac{1+|h_{m2}|^2c_{m2}^2\overline{\gamma}_{m2}}{1+B^2c_{m2}^2}>C_{th}\right)\\
{=}&{\rm P_{r}}\left(B^2{<}\frac{1-c_{m2}^2C_{th}}{C_{th}c_{m1}^2c_{m2}^2},|h_{m2}|^2{>}\frac{\frac{C_{th}{-}1}{c_{m2}^2}{+}C_{th}B^2}{\overline{\gamma}_{m2}}\right),
\end{aligned}
\end{equation}
where $\overline{\gamma}_{m2}=\frac{E_{s}}{N_{0}d_{SU_{m2}}^{\chi}}$ is the average SNR.

For notation simplicity, let $\eta=\frac{1-c_{m2}^2C_{th}}{C_{th}c_{m1}^2c_{m2}^2}$,
$\mu=\frac{C_{th}}{\overline{\gamma}_{m2}}$,
and $\nu=\frac{(C_{th}-1)}{c_{m2}^2\overline{\gamma}_{m2}}$. Note that
$\eta$ must be greater than zero, otherwise, ${\rm SOP}=1$. Since
$\eta=\frac{1-c_{m2}^2C_{th}}{C_{th}c_{m1}^2c_{m2}^2}>0$,
we can obtain $c_{m2}^2C_{th}<1$. Finally, ${\rm P_{r1}}$ can be
further expressed as
\begin{equation}
\begin{aligned}
{\rm P_{r1}}<&{\rm P_r}(B^2<\eta,|h_{m2}|^2>\mu B^2+\nu)\\
=&\int_{0}^{\eta}\int_{\mu x+\nu}^{\infty}{\rm exp}(-y)dy\frac{1}{\lambda_E}{\rm exp}\left(-\frac{x}{\lambda_E}\right)dx\\
=&\frac{e^{-\nu}}{\mu \lambda_E+1}[1-e^{-(\mu+\frac{1}{\lambda_E})\eta}].
\end{aligned}
\end{equation}

Therefore, with (14)-(16), the SOP for the group $m$ can be given by
\begin{equation}
\begin{aligned}
{\rm P_r}(\min\{C_{m1},C_{m2}\}<R)=1-\frac{e^{-\nu}}{\mu \lambda_E+1}[1-e^{-(\mu+\frac{1}{\lambda_E})\eta}].
\end{aligned}
\end{equation}

Finally, with (13) and (17), the SOP of the whole system can be written as
\begin{equation}
\begin{aligned}
{\rm SOP}=\left[1-\frac{e^{-\nu}}{\mu \lambda_E+1}[1-e^{-(\mu+\frac{1}{\lambda_E})\eta}]\right]^{M}.
\end{aligned}
\end{equation}

\subsection{Asymptotic SOP analysis}\label{subsec3.2}
The above analytical result is related to $\mu$, $\nu$, and $\lambda_E$, which can not provide an explicit insight. Thus, we
provide an asymptotic analysis. In particular, $\mu$ and $\nu$ become
zero when $\overline{\gamma}_{m2}\rightarrow\infty$. Then, the SOP of
the group $m$ can be asymptotically written as
\begin{equation}
\begin{aligned}
{\rm P_r}(\min\{C_{m1},C_{m2}\}<R)&\approx1- {\rm P_r}\left(\frac{1+\frac{c_{m1}^2}{c_{m2}^2}}{1+B^2c_{m1}^2}>C_{th}\right)\\
&={\rm exp}\left(-\frac{\eta}{\lambda_E}\right).
\end{aligned}
\end{equation}

With (13) and (19), the system SOP in a multi-user scenario can be asymptotically expressed as
\begin{equation}
{\rm SOP}\approx{\rm exp}\left[\left(-\frac{\eta}{\lambda_E}\right)\right]^{M}.
\end{equation}

Above expression indicates that the asymptotic SOP is only related
to $\lambda_E$ and $M$, and it tends to a constant when
$\overline{\gamma}_{m2}\rightarrow\infty$. Thus, at high SNRs, the secrecy performance is only related to the quality of
eavesdropping link and $N$. Interestingly, increasing $N$ results
in poor secrecy performance, but increasing $M$ results in better secrecy performance.
\section{Numerical results}\label{sec4}
\begin{figure}[H]
\centering
\includegraphics[width=8cm,height=6cm]{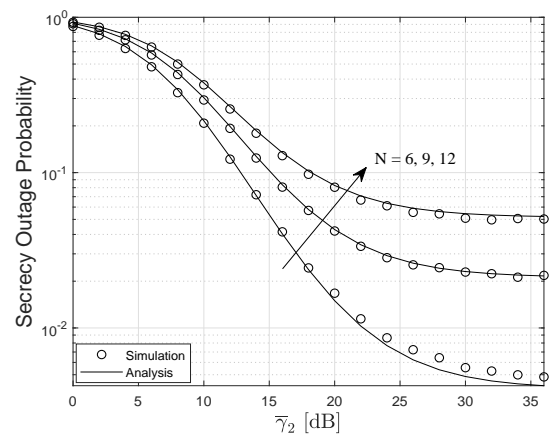}
\caption{SOP for different numbers of intelligent surfaces. $R$ =\ 0.05 bit per channel use (BPCU), $c_{m1}^2 = 0.95$, $M$ = 2, $\overline{\gamma}_{E} = 0\ {\rm dB}$.}
\end{figure}
In this section, some numerical results are provided to illustrate
the secrecy performance of our proposed NOMA system. Meanwhile,
Monte-Carlo simulation results are provided to verify our
analysis. Without loss of generality, we assume that
$\overline{\gamma}_{m2}$ = $\overline{\gamma}_{2}$ and $\overline{\gamma}_{SR_{m}E}$ = $\overline{\gamma}_{SE}$ = $\overline{\gamma}_{E}$ for all $m$.

In Fig.2, we plot the SOP curves for different $N$ when $M=2$. We can see
that our theoretical calculation and simulation are consistent.
From Fig.2, it is shown that $N$ has a great impact on the system
performance. The SOP becomes higher when
$N$ increases. The reason is that although the RIS does not adjust
the phase for the RIS-E channel to maximize $\gamma_{E_{m1}}$, E still
receives $N$ copies of the signals from the RIS. Thus, E also
enjoys the advantage induced by the RIS. For large
$\overline{\gamma}_{2}$, from Eq.(4), we can see that
$\gamma_{m1}=c_{m1}^2/c_{m2}^2$ is a constant. Thus, large $N$ results in a
large $\gamma_{E_{m1}}$ and in turn results in a higher SOP since
${\rm P_r}(\min\{C_{m1},C_{m2}\}<R)$.
\begin{figure}[H]
\centering
\includegraphics[width=8cm,height=6cm]{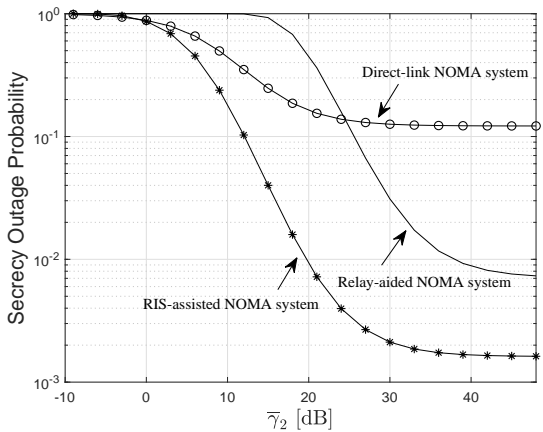}
\caption{SOP for different NOMA schemes. $N$ = 5, $M$ = 2, $\overline{\gamma}_{E}=0\ {\rm dB}$.}
\end{figure}
\begin{figure}[H]
\centering
\includegraphics[width=8cm,height=6cm]{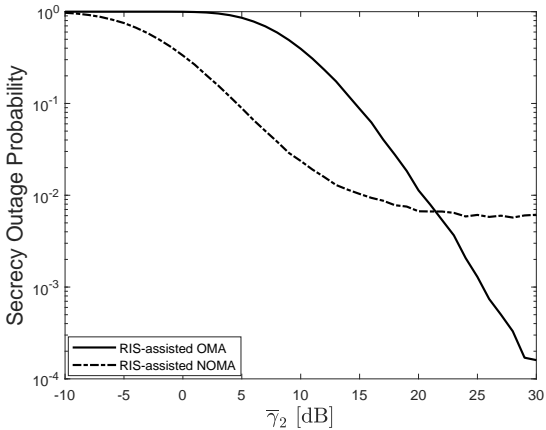}
\caption{SOP comparison between RIS-assisted NOMA systems and RIS-assisted OMA systems. $R$ =\ 0.3 BPCU, $N$ = 5, $M$ = 2, $\overline{\gamma}_{E}=0\ {\rm dB}$.}
\end{figure}
In Fig.3, we present a SOP comparison between different NOMA schemes. It is clearly observed that the system performance by
using RISs is significantly improved compared to the direct-link
NOMA system and the relay-aided NOMA system. In Fig.4, we plot a SOP comparison between the RIS-assisted NOMA system and the RIS-assisted OMA system. At low SNRs, the RIS-aided NOMA system has a better system performance than the RIS-aided OMA system. At high SNRs, the interference in NOMA users become dominant, which affects the system performance. However, OMA system has no interference between users, which in turn results in a good performance at high SNRs.
\begin{figure}[H]
\centering
\includegraphics[width=8cm,height=6cm]{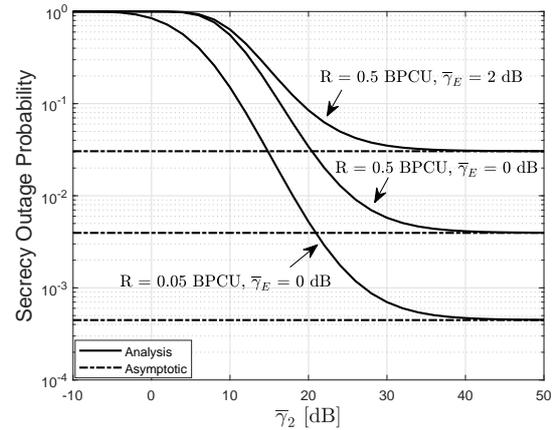}
\caption{SOP for different values of $\overline{\gamma}_{E}$ and target rates. $N$ = 4, $M$ = 2.}
\end{figure}
In Fig.5, we plot the SOP curves versus $\overline{\gamma}_2$ for different average
SNRs of the wiretap link and target rates. In Fig.6, we plot the SOP curves for different $M$.
From Fig.5 and Fig.6, it is demonstrated that the SOP tends to a constant for large
$\overline{\gamma}_{2}$, which verifies our asymptotic analysis in
Section 3.2. Also, we can see that large $M$ can improve the system performance.
\begin{figure}[H]
\centering
\includegraphics[width=8cm,height=6cm]{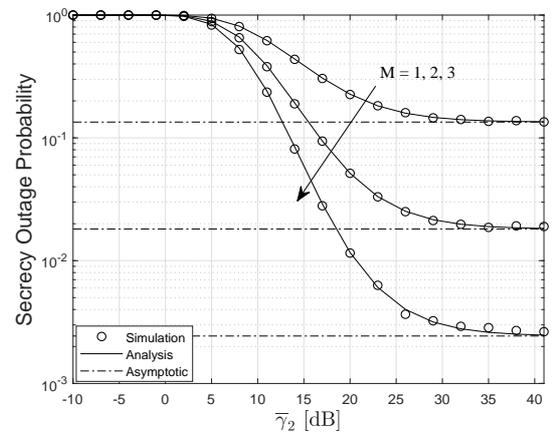}
\caption{SOP for different numbers of paired groups $M$. $R$ =\ 0.3 BPCU, $N$ = 7, $\overline{\gamma}_{E}=0\ {\rm dB}$.}
\end{figure}
\section{Conclusions}\label{sec5}
In this paper, we analyzed the SOP of  RIS-assisted NOMA systems.
Results reveal that SOP tends to a constant at high SNRs.
Moreover, increasing the number of intelligent
elements has a negative impact on the system secrecy performance
since E also takes advantages of the RIS. However, the secrecy performance can be improved by using the group selection.

\section{References}\label{sec6}

\begin{enumerate}

\item[{[1]}] Ding, Z., Liu, Y., Choi, J.,~et al.:'Application of non-orthogonal multiple access in LTE and 5G networks, IEEE Commun. Mag., 2017, 55, (2), pp 185-191

\item[{[2]}] Chen, J., Yang, L., Alouini, M.:'Performance analysis of cooperative NOMA schemes in spatially random relaying networks, IEEE Access, 2018, 6, pp 33159-33168

\item[{[3]}] Lei, H., Zhang, J., Park, K.H.,~et al.:'Secrecy outage of max-min TAS scheme in MIMO-NOMA systems, IEEE Trans. Veh. Technol., 2018, 67, (8), pp 6981-6990

\item[{[4]}] Liu, Y., Pan, G., Zhang, H., Song, M.:'Hybrid decode-forword and amplify-forward relaying with non-orthogonal multiple access, IEEE Access, 2016, 4, pp 4912-4921

\item[{[5]}] Jiao, R., Dai, L., Zhang, J., MacKenzie, R., Hao, M.:'On the performance of NOMA-based cooperative relaying systems over rician fading
channels, IEEE Trans. Veh. Technol., 2017, 66, (12), pp 11409-11413

\item[{[6]}] Yang, L., Liu, T., Chen, J., Alouini, M.:'Physical layer security for mixed $\eta$-$\mu$ and $\mathcal {M}$-distribution dual-hop RF/FSO systems, IEEE Trans. Veh. Technol., 2018, 67, (12), pp 12427-12431

\item[{[7]}] Yang, L., Hasna, M., Ansari, I.:'Physical layer security for TAS/MRC systems with and without co-channel interference over $\eta$-$\mu$ fading channels, IEEE Trans. Veh. Technol., 2018, 67, (12), pp 12421-12426

\item[{[8]}] Xiang, Z., Yang, W., Pan, G., Cai, Y., Song, Y.:'Physical layer security in cognitive radio inspired NOMA network, IEEE J. Sel. Topics Signal Process., 2019, 13, (3), pp 700-714

\item[{[9]}] Zhao, X., Chen, H., Sun, J.:'On physical-layer security in multiuser visible light communication systems with non-orthogonal multiple access, IEEE Access, 2018, 6, pp 34004-34017

\item[{[10]}] Chen, J., Yang, L., Alouini, M.:'Physical layer security for cooperative NOMA systems, IEEE Trans. Veh. Technol., 2018, 67, (5), pp 4645-4649

\item[{[11]}] Basar, E., Renzo, M., Rosny, J., Debbah, M., Alouini, M., Zhang, R.:'Wireless communications through reconfigurable intelligent surfaces, IEEE Access, 2019, 7, pp 116753-116773

\item[{[12]}] Yang, L., Guo, W., Ansari, I.S.:'Mixed dual-hop FSO-RF communication systems through reconfigurable intelligent surface, IEEE Commun. Lett., 2020, 24, (7), pp 1558-1562

\item[{[13]}] Yang, L., Meng, F., Zhang, J., Hasna, M.O., Tsiftsis, T., Renzo, M.D.:'On the performance of RIS-assisted dual-hop UAV communication systems, IEEE
Trans. Veh. Technol., Early Access, DOI:10.1109/TVT.2020.3004598

\item[{[14]}] Yang, L., Yang, Y., Hasna, M.O., Renzo, M.D.:'Coverage, probability of SNR gain, and DOR analysis of RIS-aided communication systems, IEEE Wireless Commun. Lett., Early Access, DOI: 10.1109/LWC.2020.2987798

\item[{[15]}] Huang, C., Zappone, A., Alexandropoulos, G.C., Debbah, M., Yuen, C.:'Reconfigurable intelligent surfaces for energy efficiency in wireless communication, IEEE Trans. Wireless Commun., 2019, 18, (8), pp 4157-4170

\item[{[16]}] Huang, C., Alexandropoulos, G.C., Yuen, C., Debbah, M.:'Indoor
signal focusing with deep learning designed reconfigurable intelligent surfaces, Proc. IEEE IWSPAWC, Cannes, France, 2019, pp. 1-5

\item[{[17]}] Fu, M., Zhou, Y., Shi, Y., Letaief K.:'Reconfigurable intelligent surface empowered
downlink non-orthogonal multiple access, [Online]. Available: https://arxiv.org/abs/1910.07361

\item[{[18]}] Hou, T., Liu, Y., Song, Z., Sun, X., Chen, Y., Hanzo, L.:'Reconfigurable intelligent surface aided
NOMA networks, [Online]. Available: https://arxiv.org/abs/1912.10044

\item[{[19]}] Zhu, J., Huang, Y., Wang, J., Navaie, K., Ding, Z.:'Power efficient IRS-assisted NOMA, [Online]. Available: https://arxiv.org/abs/1912.11768

\item[{[20]}] Zheng, B., Wu, Q., Zhang, R.:'Intelligent reflecting surface-assisted multiple access with
user pairing: NOMA or OMA?, [Online]. Available: https://arxiv.org/abs/2001.08909

\item[{[21]}] Thirumavalavan, V., Jayaraman, T.:'BER analysis of reconfigurable intelligent surface
assisted downlink power domain NOMA system, [Online]. Available: https://arxiv.org/abs/2002.09453

\item[{[22]}] Cheng, Y., Li, K., Liu, Y., Teh, K., Poor, H.:'Downlink and uplink intelligent reflecting
surface aided networks: NOMA and OMA, [Online]. Available: https://arxiv.org/abs/2005.00996

\item[{[23]}] Ding, Z., Poor, H.:'A simple design of IRS-NOMA transmission, IEEE Commun. Lett., 2020, 24, (5), pp 1119-1123

\item[{[24]}] Ding, Z., Fan, P., Poor, H.:'Impact of user pairing on 5G nonorthogonal
multiple-access downlink transmissions, IEEE Trans. Veh. Technol., 2016, 65, (8), pp 6010-6023

\item[{[25]}] Liu, Y., Qin, Z., Elkashlan, M., Gao, Y., Hanzo, L.:'Enhancing the physical layer security of non-orthogonal multiple
access in large-scale networks, IEEE Trans. Wireless Commun., 2017, 16, (3), pp 1656-1672

\item[{[26]}] Proakis, J.G., Salehi, M.:'Digital Communications Fifth Edition, McGraw-Hill Education Press, 2008, 45-46

\item[{[27]}] Yang, L., Yang, J., Xie, W., Hasna, M., Tsiftsis, T., Renzo, M.:'Secrecy performance analysis of RIS-aided wireless communication systems, IEEE Trans. Veh. Technol., Early Access, DOI: 10.1109/TVT.2020.3007521

\end{enumerate}

\end{document}